# Generative AI and Power Imbalances in Global Education: Frameworks for Bias Mitigation

**Matthew Nyaaba[1,2], Alyson Wright[2], and Gyu Lim Choi[2]**

[1]AI4STEMEducation Center, University of Georgia, Athens, GA 30602 USA.

[2]Department of Educational Theory and Practice, University of Georgia, Athens, GA 30602 USA.

**Abstract**

This study interrogates how Generative Artificial Intelligence (GenAI) reproduces power imbalances hierarchies in education and advances an evidence-based framework to mitigate resulting inequities. Employing a Critical Constructive Qualitative design, the study conducted zero-shot prompt testing using leading systems, ChatGPT-4 Turbo and Gemini 1.5, and documented outputs through real-time screenshots across Global North and South contexts. A critical interpretive analysis traced textual, visual, and structural patterns indicative of digital neocolonialism and their implications for educational equity and transformation. Findings reveal six interconnected dimensions through which GenAI sustains Western hegemony while also holding potential to reshape educational practice. Western curriculum ideologies were evident when Gemini was asked to name the seasons in both the United States and Ghana, yet produced identical four-season responses, reflecting Western climatological assumptions and overlooking regional realities. This illustrates how algorithmic bias can distort learners' understanding of the world and perpetuate Global North knowledge dominance. Other outputs revealed cultural imperialism through stereotypical imagery, pedagogical control through Western-centered lesson examples, language marginalization due to limited support for Indigenous and local languages, racial and ethnic underrepresentation in visual outputs reflecting predominantly Western identities, and digital divide dynamics stemming from subscription-based models that privilege resource-rich users and regions. These dimensions demonstrate how GenAI can entrench educational inequities even as it offers opportunities for pedagogical innovation. Synthesizing these results, we propose a dual-pathway mitigation model to advance equitable AI in education. The *Inclusive AI Design* framework includes (1) Liberatory Design Methods to center non-Western epistemologies, (2) Foresight by Design to anticipate harm and ensure representational equity, and (3) Decentralizing GenAI hubs to incorporate local voices and data sovereignty. The pedagogical-level pathway, *Human-Centric Prompt Engineering*, equips educators to contextualize prompts and critically engage outputs. These pathways position GenAI as a transformative force toward equitable and culturally responsive education.

## Introduction

The emergence of generative artificial intelligence (GenAI) technologies represents a significant advancement in the digital space, transforming how data is utilized and how content is generated (Ipek et al. 2023). GenAI encompasses a range of technological capabilities, producing complex outputs such as text, audio, images, and now, video, based on patterns learned from vast datasets (Lim et al., 2023). This ability positions GenAI as a tool of immense potential within various sectors, including education, where it is revolutionizing teaching, learning, assessment, and research (Latif et al., 2023; Zhai et al., 2024). GenAI is currently noted for enhancing personalized learning experiences, automating scoring and grading, and aiding in lesson preparation. As educational institutions, from K-12 to higher education, continue to integrate GenAI into their practices, it becomes crucial to understand the foundational mechanisms, challenges, and biases associated with these technologies to harness their capabilities effectively (Ipek et al., 2023; Latif et al., 2023).

While the benefits of GenAI in education are substantial, numerous studies have identified accompanying challenges, particularly the inherent biases within GenAI systems (Arora et al., 2023). These biases largely stem from the fact that GenAI systems learn from existing data, thereby absorbing and replicating the biases present within that data (Lynch et al., 2023; Van Niekerk et al., 2024). The presence of these biases can potentially lead to skewed knowledge dissemination, favoring dominant cultural narratives, typically Western, while overlooking minority perspectives and non-Western narratives (Lynch et al., 2023). The consequences of this could foster traditional colonial ideologies, amplify existing educational disparities, and perpetuate digital neocolonialism.

Digital neocolonialism in the domain of GenAI specifically describes a scenario where control over GenAI technologies and the data they use is concentrated in the hands of a few, predominantly Western, corporations or countries (Adam, 2019; Menon, 2023; Sok & Heng, 2023). This concentration of control can impose a digital hegemony on less technologically advanced nations, echoing the power imbalances seen in traditional colonial relationships (Zembylas, 2023), where economic and cultural dominance was asserted by colonial powers over colonized regions. Acknowledging these biases and the potential for digital neocolonialism is crucial to ensure that GenAI in education serves global needs equitably (Zembylas, 2023). The study unfolded in two key phases. In the first, a critical phase, zero-shot prompt testing was conducted using ChatGPT-4 Turbo and Gemini 1.5 to examine how GenAI systems represent educational contexts across the Global North and South. Outputs were documented through real-time screenshots and analyzed interpretively to trace textual, visual, and structural indicators of digital neocolonialism. In the second, a constructive phase, findings from this analysis informed the creation of the *Inclusive AI Design Framework* and *Human-Centric Prompt Engineering Model*. We were guided by the following questions:

1. How do Generative AI outputs reflect global and cultural biases within educational contexts?
2. How can cultural biases present in Generative AI be mitigated for global education?

## Positionality Statement

The authors of this study represent diverse cultural, academic, and professional backgrounds spanning Africa, Asia, and North America. All are early-career scholars with active research and teaching engagements in K–12 and higher education contexts, particularly in the areas of early childhood, elementary, middle, teacher education, and with research focus on AI in education, culturally responsive pedagogy, and educational equity. The first author led the overall study of design, critical analysis, and development of the conceptual frameworks, drawing on his experiences in the Global South. The second author contributed to the mitigation approaches, drawing from her teaching experiences and technological exposure to educational technologies in the Global North. The third author provided contextual validation and cultural interpretation, ensuring that the findings reflect perspectives across multiple regions. The authors continuously engaged in collaborative discussions, intercultural reflection, and iterative consensus-building to verify interpretations and minimize positional bias. This reflexive process helped the team remain critically aware of how their own experiences informed their interpretation of AI outputs, while ensuring that analyses were balanced, evidence-based, and globally contextualized.

## Literature Review

### Digital Neocolonialism

Digital neocolonialism refers to the perpetuation of historical colonial inequities through the use of digital technologies, illustrating the exploitation and control exerted by dominant nations or large technology companies over nondominant countries (Gravett, 2022). This dynamic has led to new global power imbalances, with large technology companies accumulating massive amounts of data for profit (Gerbrandy & Phoa, 2022). The neocolonial project in the digital age operates through platforms like social media and technologies, including GenAI, influencing democratic processes and perpetuating power imbalances (Zembylas, 2023).

Digital neocolonialism manifests in various ways, such as reinforcing inequalities, generalizing language and culture, controlling curriculum and content, and dominating the market and economy. For example, major corporations often control and exploit data, extracting significant value while providing little benefit to the individuals from whom the data is collected (Coleman, 2018; Stürmer et al., 2021). Additionally, the dependence of developing countries on technology infrastructure and platforms provided by foreign companies limits their ability to develop and control their digital ecosystems, hindering local industry development and innovation (Adam, 2019; Kwet, 2019; Langmia & Sani, 2023).

The global spread of dominant cultural norms through digital platforms can lead to cultural imperialism, marginalizing local cultures and languages (Enein, 2023). Digital cultural hegemony undermines the diversity and autonomy of indigenous communities, as traditional forms of expression and knowledge are overshadowed by multinational tech companies. Consequently, this may lead to the homogenization of GenAI content and limit unique local perspectives (Hammer & Park, 2021; Couldry & Mejias, 2019). This dominance affects education in developing countries, as seen in Adam's (2019) study on MOOCs and digital textbooks produced by institutions in the Global North serving learners in the Global South.

Addressing digital neocolonialism is crucial for safeguarding sovereignty, privacy, and democratic values, especially in education, where GenAI is already in near full operation (Van Niekerk et al., 2024).

## Generative AI in Education

GenAI is revolutionizing education on a global scale (Lim et al., 2023). Current research on the application of GenAI in education typically falls within the three domains of teaching and learning, assessment, and research (Chiu, 2021; Ding et al., 2024; Nyaaba, 2024). Recent studies demonstrate that GenAI influences several educational domains, including personalized learning, adaptive learning, immediate feedback, lesson planning, automated assessment, and research (Latif et al., 2023).

GenAI can support personalized learning, but it may also lead to over-reliance and reduced critical thinking (Bai et al., 2023). Generally, GenAI motivates learners to develop reading and writing skills, although its impact on listening and speaking skills remains neutral (Ali et al., 2023). For instance, Li et al. (2023) explored the perspectives of YouTube content creators on using GenAI in self-directed language learning (SDLL). Their findings indicate that GenAI is valuable for its availability, versatility, and transformative potential in enhancing SDLL, providing contextually relevant responses, and fostering meaningful learner interactions (Li et al., 2024). Additionally, research by Lee and Zhai (2024) revealed that pre-service teachers (PSTs) effectively integrated ChatGPT into science learning, scoring well on a modified TPACK-based rubric, particularly in 'instructional strategies & ChatGPT.' However, they demonstrated less proficiency in using ChatGPT's full potential, indicating the need for high-quality questioning, self-directed learning, individualized support, and formative assessment to enhance lesson planning (Choi et al., 2024; Lee & Kang, 2024).

Furthermore, studies indicate that GenAI has the potential to support educators in research activities; however, inaccuracies and biases, such as language limitations and a lack of context, remain prevalent (Owoahene & Nyaaba, 2024; Nyaaba et al., 2024). A key takeaway from these studies is that GenAI has the potential to transform education, but is accompanied by biases. Consequently, students are at the forefront of GenAI usage. For example, Holland and Ciachir (2024) posited that students appreciated the immediacy and validation provided by GenAI, but concerns about equity and academic integrity emerged, particularly in group work where misuse of the tool could lead to academic misconduct affecting all members. Similarly, Rasul et al. (2023) identified academic integrity as a significant challenge in their examination of GenAI roles in higher education. They emphasized the need for clear institutional policies and transparency to mitigate these risks and ensure ethical GenAI usage. There is a strong suggestion for well-defined guidelines and policies to support the responsible and effective integration of GenAI in education (Holland & Ciachir, 2024; Rasul et al., 2023).

## Method

We employed critical constructive qualitative design grounded in the framework of *Digital Colonialism* (Freire, 2020; Kincheloe & McLaren, 2011). Guided by the *Digital*

*Colonialism/Imperialism* framework (Couldry & Mejias, 2019; Kwet, 2019), this method allowed us to critically interpret the GenAI outputs to expose how algorithmic systems replicate or resist global asymmetries in language, culture, and epistemic authority. This was then followed by a constructive proposal of mitigation frameworks designed to inform equitable and culturally responsive AI in education.

**Study Context**

This study was conducted between March and May 2024, during which two of the most advanced publicly accessible GenAI systems were employed, *ChatGPT-4 (Turbo, Advanced Mode) developed by OpenAI, and Gemini 1.5 developed by Google DeepMind*. At the time of data collection, the *Advanced Mode* of ChatGPT provided access to GPT-4 Turbo, an OpenAI's premium large-language-model variant known for its enhanced reasoning, extended context window, and faster processing speed (OpenAI, 2024). In parallel, *Gemini 1.5*, which evolved from *Bard* (launched in March 2023 and rebranded in December 2023), also offered multimodal reasoning and cross-linguistic capabilities suitable for comparative testing (Google DeepMind, 2024). The concurrent use of these two state-of-the-art models allowed for a comparative examination of how leading GenAI systems represent cultural contexts and reproduce, or resist, digitally neocolonial tendencies.

**Data Collection**

The study employed a structured empirical documentation strategy to generate verifiable evidence of GenAI's cultural representations and underlying biases. Using a zero-shot prompting technique, queries were entered without prior examples or contextual guidance to isolate each model's inherent epistemic and cultural assumptions. This approach allowed the researchers to identify baseline patterns of bias without the influence of human framing or priming effects. A series of systematically designed prompts were administered across six thematic domains, knowledge representation, visual culture, pedagogy, access, language, and race.

Each query was tested under controlled conditions, and outputs were captured through real-time screenshots that preserved the complete prompt, metadata, platform, date, and location to ensure transparency and replicability. For example, prompts such as "What seasons are there in a year?" and "Generate an image of a human robot" illustrate the types of queries used to assess GenAI's default assumptions about culture, geography, and representation. To ensure contextual robustness, identical prompts were tested across multiple cultural and geographic regions, such as Ghana and the United States, to determine whether responses varied by location. Findings revealed consistent Western-centric orientations across all contexts, underscoring the systemic nature of GenAI's cultural encoding.

**Analytical Approach**

The analysis of the generated GenAI outputs followed a critical interpretive approach, guided by the concept of digital neocolonialism (Gravett, 2022; Kwet, 2019). Each AI response and screenshot was examined for recurring patterns of cultural bias, representational imbalance, and linguistic exclusion that reflected the global power asymmetries described in the literature (Coleman, 2018; Couldry & Mejias, 2019). The process involved identifying textual and visual cues that reproduced colonial hierarchies, such as Western-centric imagery, English-dominant language use, and assumptions of Global North educational norms (Adam, 2019; Zembylas, 2023). These observed patterns were then interpreted through postcolonial and critical algorithmic lenses to determine how GenAI systems perpetuate or challenge digital imperialism and data dependency (Enein, 2023; Langmia & Sani, 2023; Stürmer et al., 2021). The theoretical insights from these studies informed the construction of the GenAI Neocolonialism

## Findings (Emergent Dimensions of GenAI Neocolonialism)

This section presents empirical evidence of GenAI's neocolonial manifestations in educational contexts through generated GenAI outputs, which systematically document how algorithmic biases reproduce colonial power structures. The analysis culminates in the GenAI Digital Neocolonialism Framework (see Figure 8), which theorizes the mechanisms through which technological systems perpetuate global inequalities in education. Through six interconnected dimensions, we demonstrate how GenAI systems function as vehicles for the reproduction of Western hegemony, creating what Mohamed et al. (2020) term 'algorithmic colonialism" in educational spaces. The six dimensions are as follows:

### Western Curriculum Ideologies

GenAI can develop educational content such as digital textbooks, learning modules, and automated tutoring systems. If these AI systems are primarily trained on datasets from economically dominant countries, there is a risk that the content they generate may not be culturally relevant or appropriate for students in different regions. This can lead to a form of cultural imperialism, where dominant cultures' viewpoints and knowledge systems overshadow local educational needs and values, marginalizing non-dominant cultures and languages, and reducing the diversity and richness of the educational experience for students worldwide. GenAI systems trained predominantly on data from Western contexts can inadvertently prioritize Western narratives, values, and ideologies (Bentley et al., 2024; Enein, 2023).

When these systems are used in educational content creation, the resulting materials may reflect a limited worldview, marginalizing non-Western perspectives and knowledge systems. For instance, we demonstrated the prioritization of Western knowledge by investigating the number of seasons in a year. Using the prompt "*What seasons are there in a year?"*, both Gemini and ChatGPT responded with four main seasons (see Figure 1). This answer oversimplifies climate variations and inaccurately reflects the reality in many regions of the world. For example, West Africa, Southeast Asia, and Central America have two main seasons, and polar regions can have just one. This example illustrates how algorithmic biases can reflect Western cultures and inaccurately provide contextual and diverse information and beliefs. As a tool for personalized learning, this shows the dangers that GenAI could have on educational content creation and curriculum. Although the prompts may be generic, we expected the responses to at least contextualize the answers or cover a wider range of seasonal variations around the world. Such biases in GenAI can marginalize non-dominant contents and undermining the richness and diversity of the global educational experience.

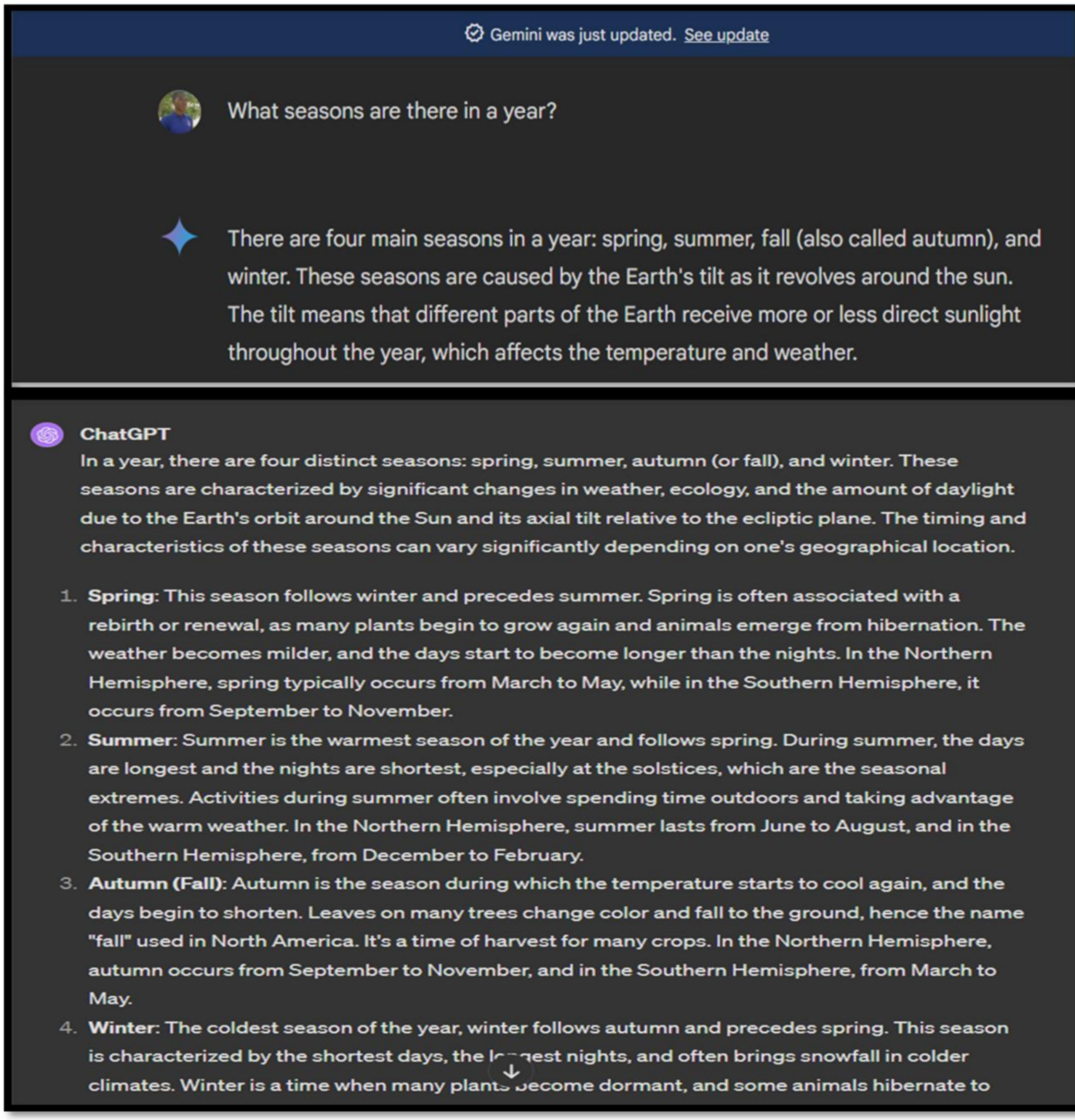

*Figure1: GenAI Reflecting Western Content Outputs*

## Cultural Imperialism

Generative AI systems might generate outputs that incorporate cultural references, examples, and case studies that might project other culture less attractive to Western cultures, as evidenced by Ożegalska-Łukasik and Łukasik's (2023)'s finding that GenAI presented a bias representation in images depicting a "*wealthy African man and his house*" versus those of a "*wealthy European man and his house*" (see Figure 2). Figure 2 illustrates how the study found that GenAI tools misrepresented what a wealthy man and his house in Africa might look like, while more accurately depicting a wealthy European man and his house. The images do not accurately represent a wealthy African man which can impact and distort students' perspectives and understandings of different cultural groups and accurate cultural contexts. A typical modern wealthy African man may live in a more sophisticated house with cars, while a traditional wealthy African man may have a large traditional house with livestock, family, and workers.

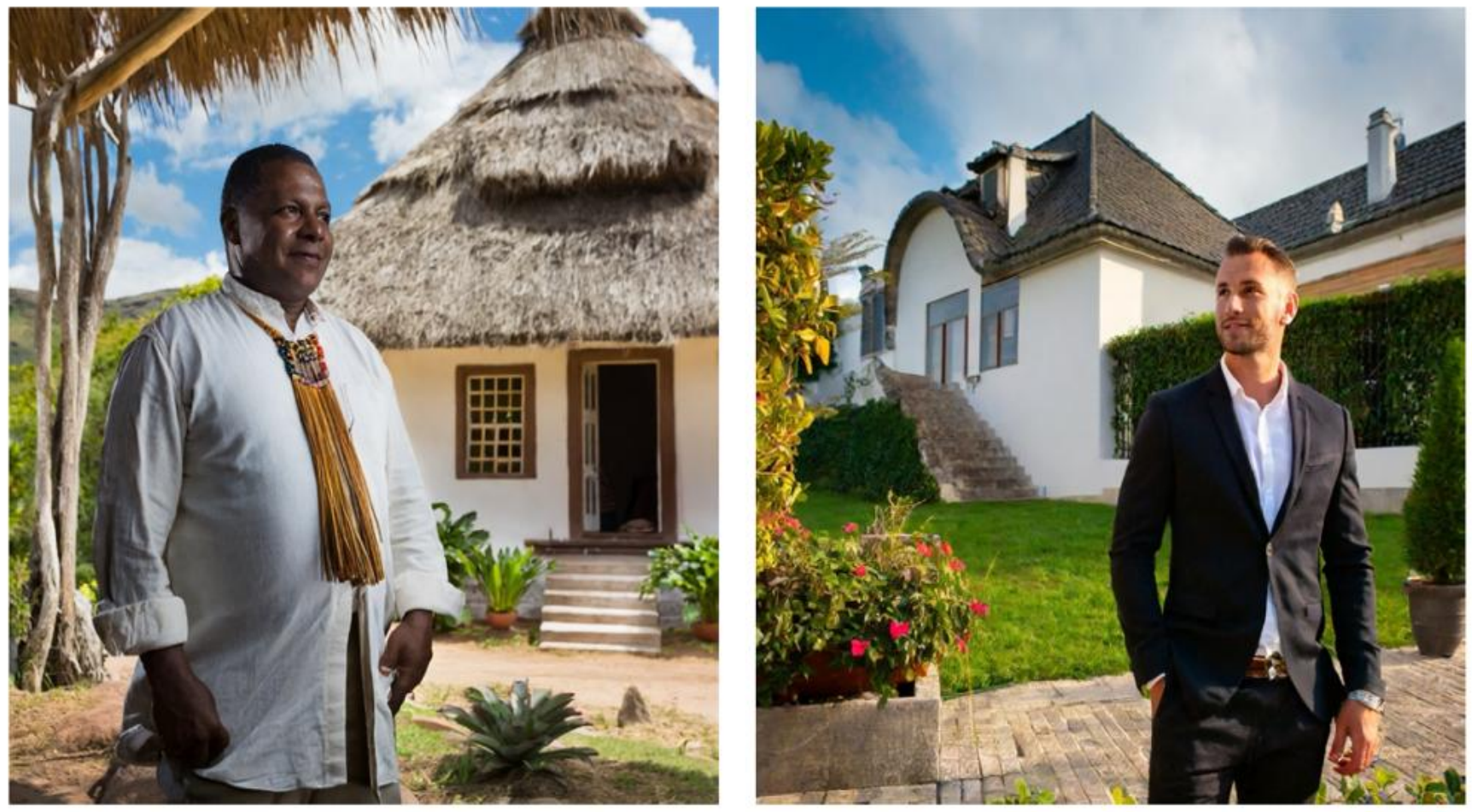

Figure 2: *An Example of AI Amplifying Cultural Bias: "Wealthy African Man and His House" vs. "Wealthy European Man and His House" Adopted from Ożegalska-Łukasik and Łukasik (2023).*

To further explore this phenomenon, we tested the findings of Ożegalska-Łukasik and Łukasik (2023) using GPT-4 and Gemini by comparing portrayals of transportation in the United States of America and Balochistan. Our study supported their observations, exemplifying the initially identified cultural bias and highlighting additional discrepancies. The responses generated by the GenAI tools depicted a stark contrast, suggesting a more advanced portrayal of the transportation system in the USA compared to the current reality of transportation in Balochistan (see Figure 3). Biased outputs can negatively influence students' understanding of various cultures and contexts, reinforcing stereotypes and misrepresentations in educational content. This cultural bias in GenAI can alienate students from other cultural backgrounds and fail to engage them with examples that resonate with their own experiences and environments (Zembylas, 2023; Athanassopoulos et al. 2023). Such cultural imperialism can lead to the marginalization of non-Western perspectives in educational content and could diminish the learning experience for students from diverse backgrounds.

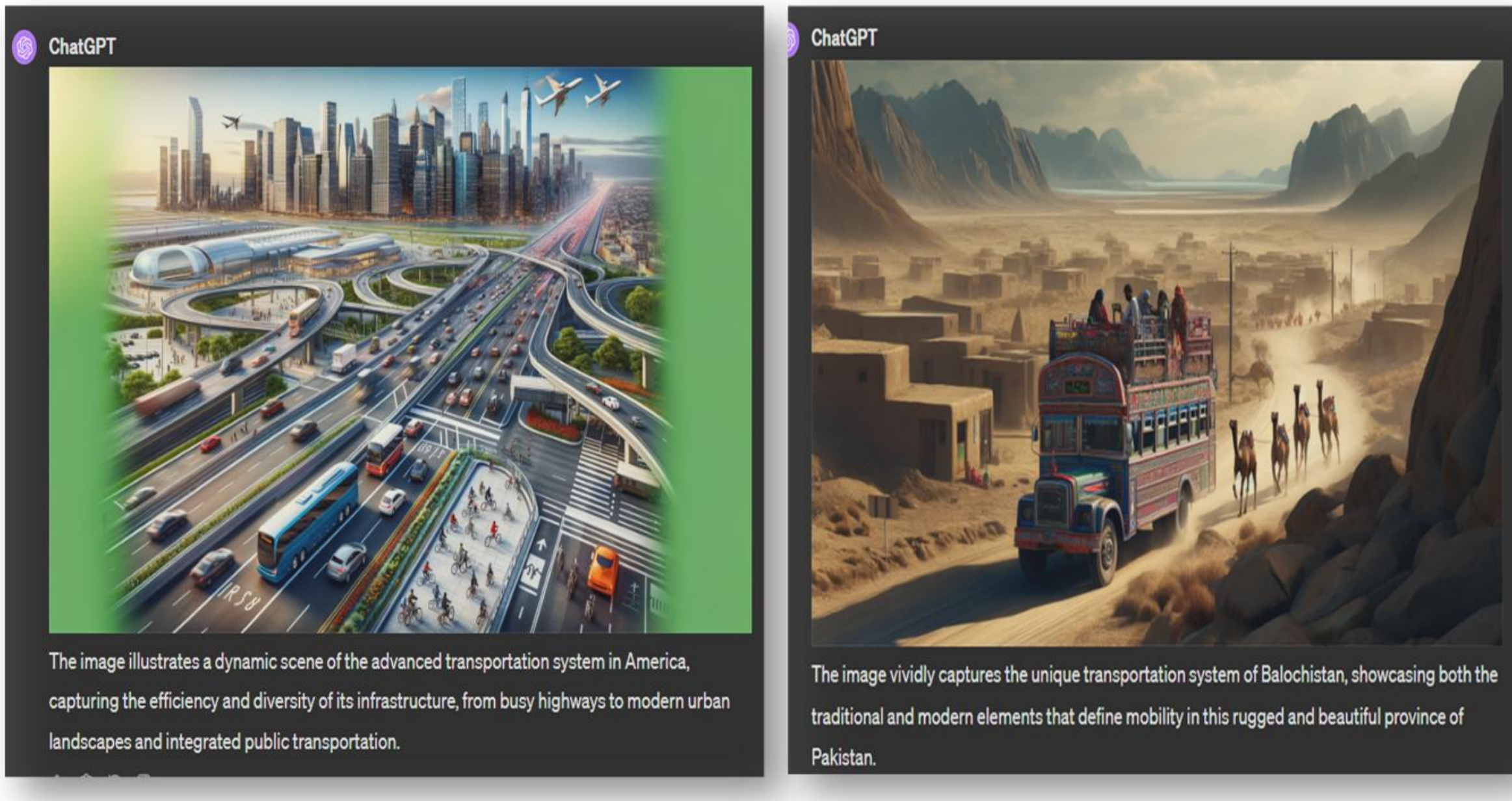

*Figure 3: Showing Cultural bias of transportation system between USA and Balochistan, a province of Pakistan.*

## Pedagogy Control

Generative AI technologies may exacerbate dominant educational philosophy and pedagogy control from Western regions than non-western regions. This might be because these technologies are primarily developed in the Western regions and therefore can reinforce ideologies and teaching principles that may not represent local perspective (Kitsara, 2022; Lynch et al., 2023). This situation not only limits the diversity of thought in the evolution of GAI but also perpetuates a class divide within societies (Bentley et al., 2024). It begins by elucidating how colonialism distorts conventional art forms, often resulting in the erasure of indigenous cultures (Zhang & Cao, 2023). We demonstrated the pedagogy control by asking ChatGPT to prepare a lesson on the *Introduction to Environmental Science* (see Figure 4). A well-organized lesson plan was generated but exhibited materials and teaching activities that are predominantly relevant and applicable in the Western regions, for instance, as part of the lesson, students were supposed to have access to individual computers and tablets which may not be available for students in most developing countries, especially those in public K-12 schools.

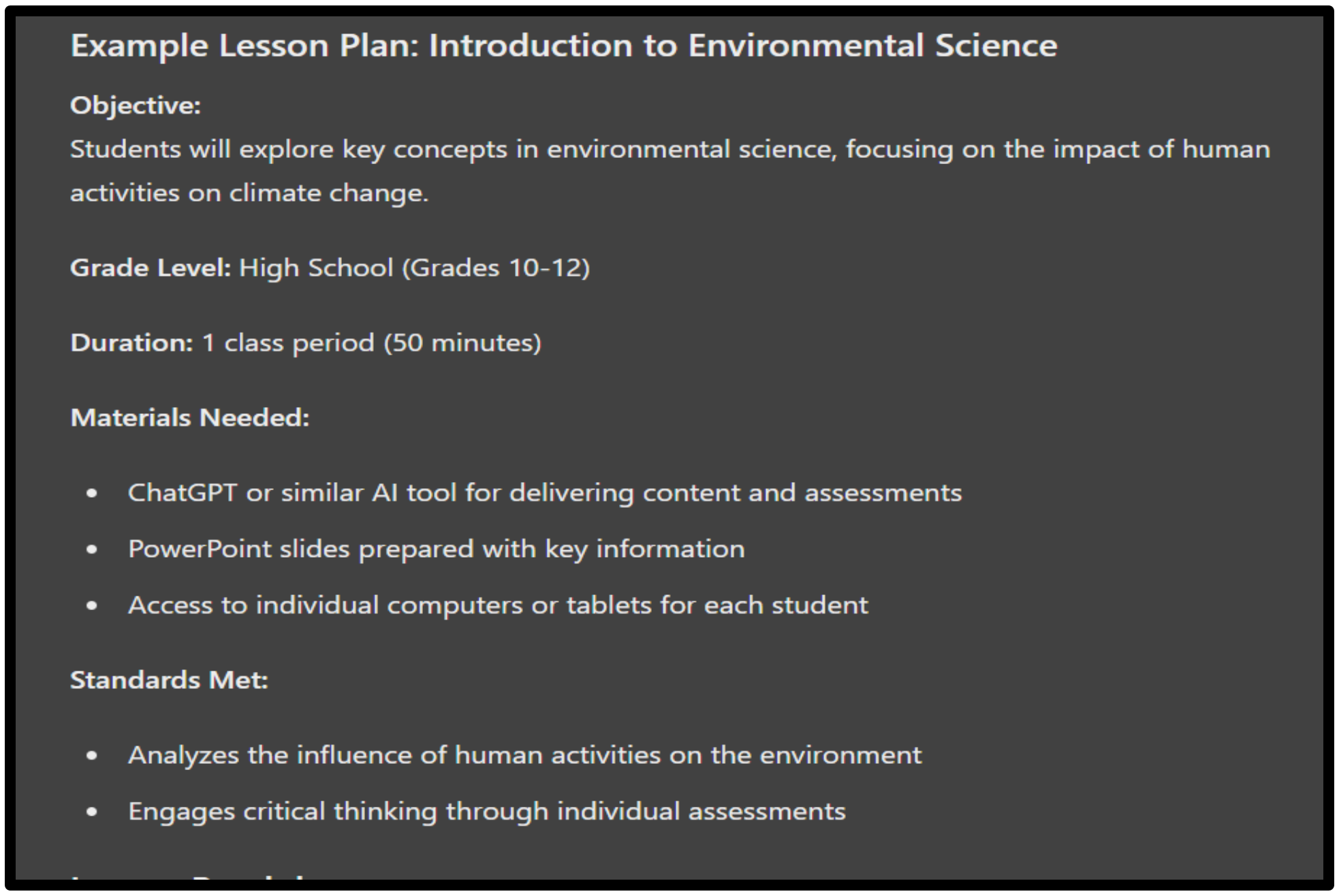

*Figure 4: Gen AI lesson plan illustrating Pedagogical Control*

**Digital Divide**

The digital divide through pricing and monthly cost is another aspect that can perpetuate neocolonialism especially, getting access to the danced versions of GenAI tools which are well-comprehensive with data and could provide users with vast features including audio and visual prompts. Gemini and GPT-4 cost over $20.00 per month which may not fall within the budget of an average teacher, parent or students in most developing countries may be privileged GenAI access to Western regions over these regions. This factor will make it difficult for both parents and schools to afford for educational purposes. For instance, Nyaaba and Zhai (2024b) study shows that teacher educators in developing countries are excited about using GenAI in their classroom practices but are much more concerned about the cost involved. In addition, Enriquez et al. (2023) posited that the monetization of GenAI tools exacerbate divide, as wealthy families can afford advanced versions of these technologies (see Figure 5), providing their children with superior educational tools and opportunities while less affluent families may be relegated to inferior, free versions of these tools, potentially widening the achievement gap between socioeconomic classes (Enein, 2023). Additionally, the restricted domain of GenAI might not be accessible to certain areas or localities. There are yet some communities who do not have access technology including GenAI due to domain and internet restrictions (Ragnedda & Ragnedda, 2020).

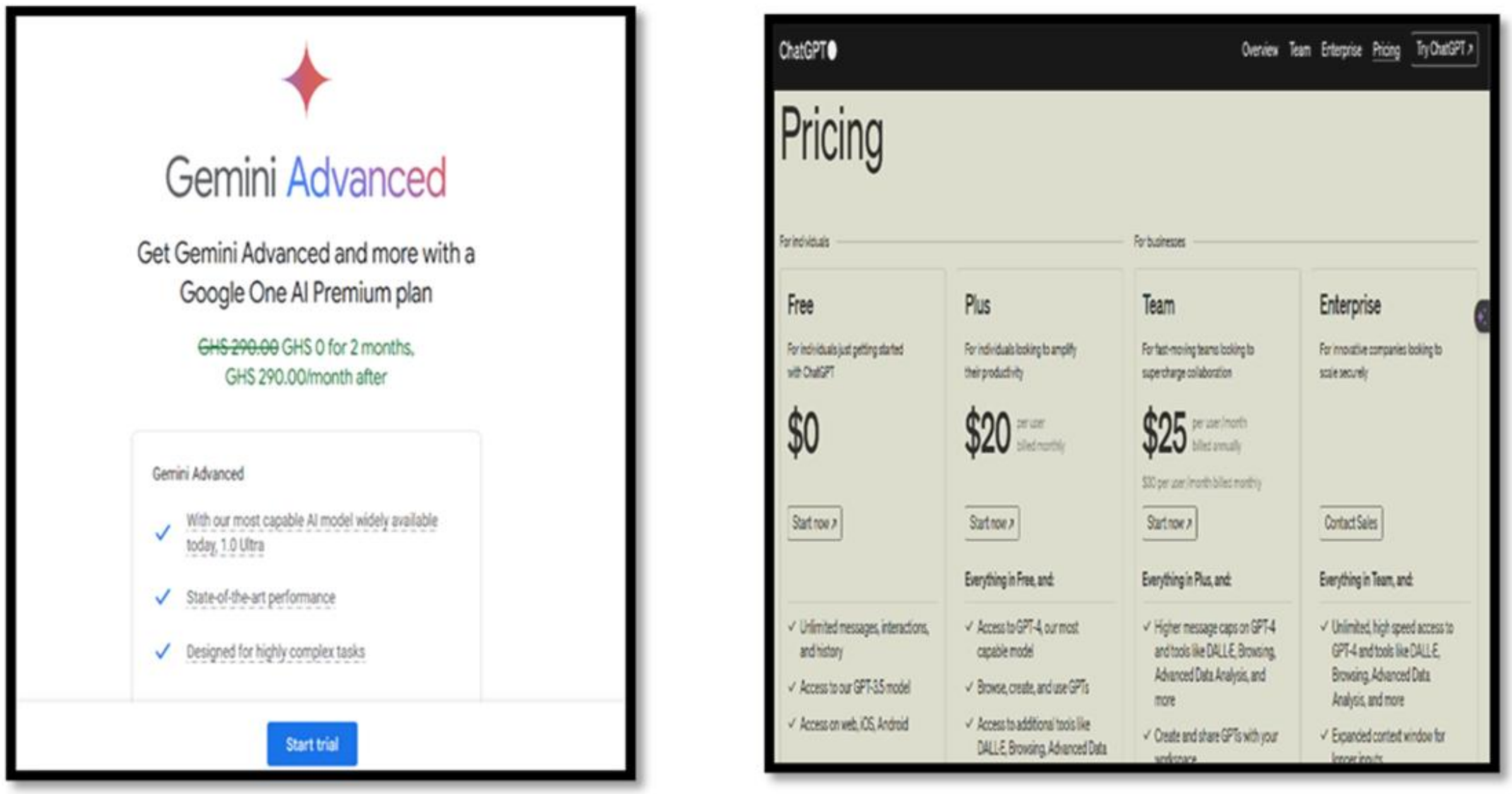

*Figure 5: Digital Divide through Pricing of GenAI tools*

**Language Marginalization**

While GenAI has significantly advanced in incorporating and translating as many languages as possible, it remains evident that numerous indigenous languages are yet to be represented within these models (Ożegalska-Łukasik & Łukasik, 2023). This may inadvertently marginalize indigenous and minority languages, potentially accelerating language erosion and the dilution of cultural heritage among certain minoritized groups. More specifically, this focus can marginalize non-dominant languages and cultures, making educational content less accessible to speakers of indigenous or less widely spoken languages. To illustrate this bias, we prompted GenAI tools to provide an alternative term for "God is Good" in Gurune (a language from Upper East, Ghana) (see Figure 6). The response showed that GenAI is indeed limited in this language and could not find the term weather in Brahui language. This may impact the ability of students to learn in their mother tongue, which is crucial for young learners' understanding and cognitive development. Educators and students are thus urged to critically evaluate the implications of using GenAI on these languages and their corresponding cultures (Share et al., 2019).

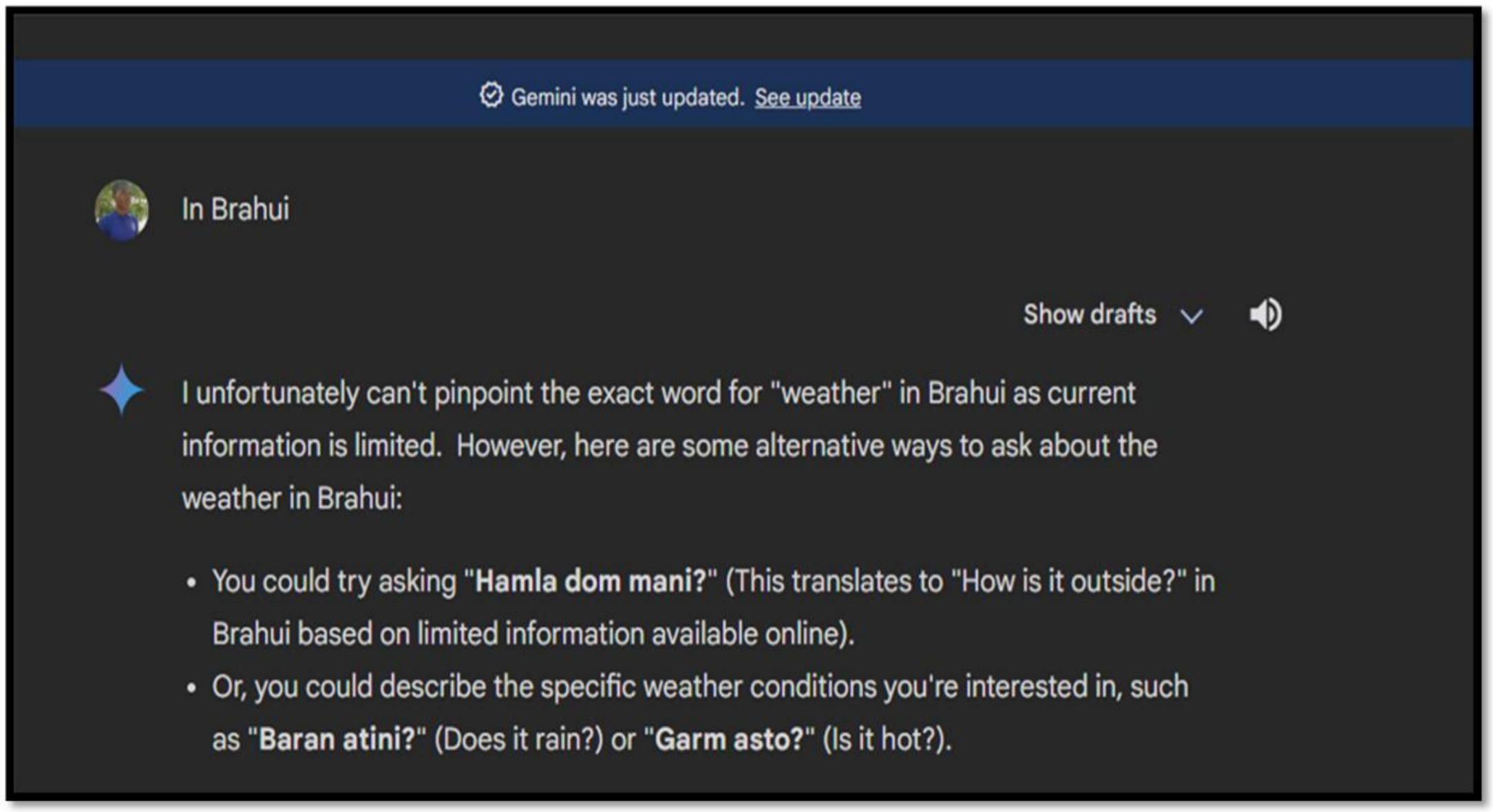

*Figure 6: GAI marginalizes indigenous and minority languages*

**Racial and Ethnic Underrepresentation**

The racial dimensions of GenAI neocolonialism manifest through what Noble's (2019) terms *Algorithms of Oppression: How Search Engines Reinforce Racism* explores how search engines perpetuate racism, such as returning pornographic results for searches of "Latinas" and primarily displaying White men in professional roles, thus perpetuating racial stereotypes (Furuhata (Furuhata, 2022; Rahman, 2020). This situation highlights the need for rectifying racial biases within search engine algorithms for fairer representations in education. We demonstrated racial biases in GenAI by requesting GenAI tools to generate an image of a "human robot," as shown in Figure 7. The output aligned with Cave and Dihal (2020)'s findings, which highlighted a white-centric bias in AI portrayals within Western culture (See figure 7). This shows the need for more inclusive representations of GenAI technologies, particularly in educational contexts. While students may be relying on GenAI tools for drawing and image generation, these tools might be representing only a group of people.

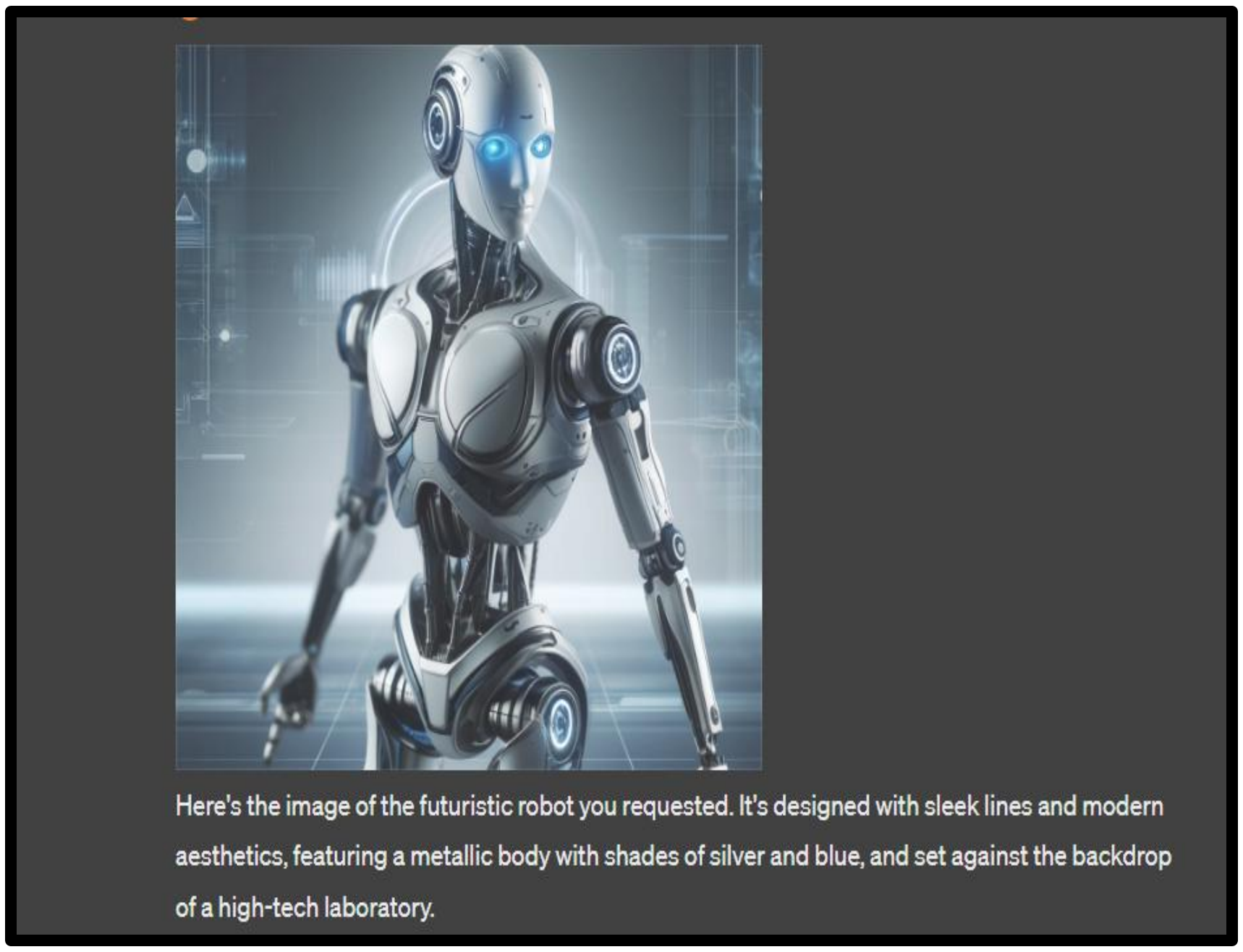

*Figure 7: GenAI exemplifying whiteness of output*

**Generative AI Digital Neocolonialism Framework**

The empirical evidence presented converges in our *GenAI and Digital Neocolonialism* Framework (see Figure 8), which theorizes how these six (6) merging dimensions operate through a central mechanism, the reproduction of global power imbalances in education. The framework positions "Reproduction of Global Power Imbalances" as the central mediating process, demonstrating how technological systems translate historical inequalities into algorithmic operations. Each dimension manifests through specific educational outcomes, dominant frameworks, excluded languages, Western ideologies, cultural misrepresentation, unequal access, and racial bias, that collectively constitute a comprehensive system of digital neocolonialism. We believe that the *GenAI and Digital Neocolonialism framework* provides a theoretical foundation for understanding these dynamics while pointing toward the necessity of transformative interventions that challenge the fundamental assumptions underlying GenAI development and implementation. Following this we proposed the mitigation framework to enhance democratization of GenAI for education in global and culturally responsive use.

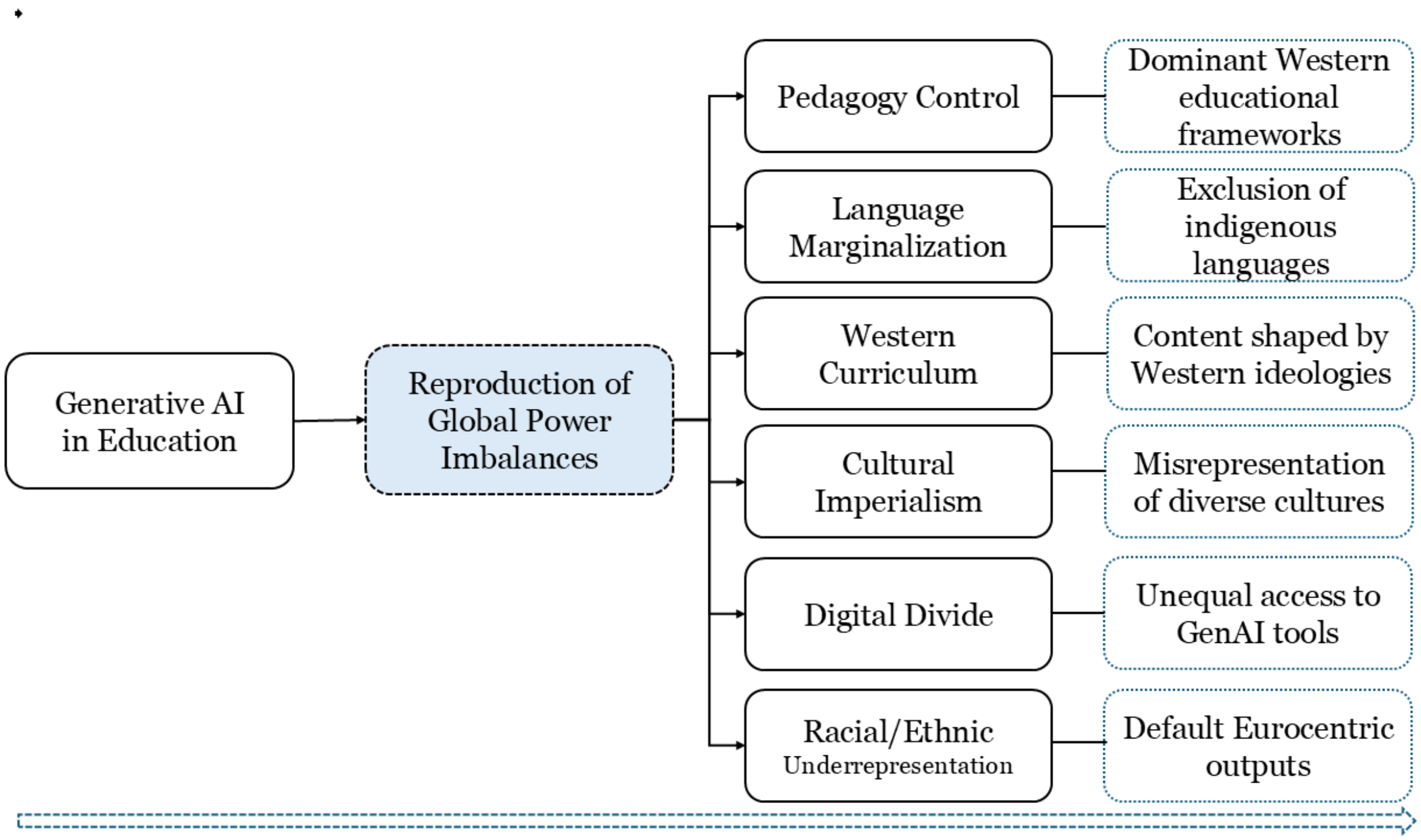

*Figure 8: GenAI Digital Neocolonialism in Education*

**Mitigation Frameworks**

Through a critical examination of GenAI and its potential to perpetuate digital neocolonialism, this section delineates comprehensive mitigation strategies operating at two distinct yet complementary levels (Brand, 2023): AI development and the pedagogical level of educational mitigations. The mitigation frameworks emerge from critical engagement with UNESCO's Recommendation on Ethics of Artificial Intelligence (UNESCO, 2021) and the AI Convention by the Council of Europe (2022), which emphasize the imperative to align AI technologies with human rights while safeguarding cultural diversity. As UNESCO (2021) articulates, addressing ethical concerns must foster innovation that upholds fundamental freedoms, while the Council of Europe (2022) advocates for frameworks that protect cultural pluralism within digital technologies. These international instruments provide the normative foundation for the dual-pathway approach delineated below. The relationship between Figures 9 and 10 reveals the complementary nature of systemic and pedagogical interventions. We proposed the *Inclusive AI Design Framework* (Figure 9), which integrates the three critical approaches targeting AI developers and designers to disrupt neocolonial patterns at their source: (1) Liberatory Design Methods to center non-Western epistemologies, (2) Foresight by Design to anticipate harm and ensure representational equity, and (3) Decentralizing GenAI hubs to incorporate local voices and data sovereignty. We proceeded to propose the *Human-Centric Prompt Engineering framework* (Figure 10) which empowers educators (users) to navigate and resist biases within existing systems. These dual pathways create a comprehensive mitigation strategy that operates across the entire GenAI ecosystem, from initial conception through classroom implementation.

## Inclusive AI Design Framework

### *Liberatory Design Methods (LDM)*

Liberatory Design Methods (LDM) ensure the inclusion of diverse perspectives and experiences. LDM empowers designers to embrace non-Western viewpoints and develop solutions by acknowledging "the intricacies of marginalized identities as a catalyst for positive innovation" (Harrington & Piper, 2018). Central to this approach is recognizing that technological solutions devised within one culture may not seamlessly translate to another. Liberatory design amplifies the voices of underrepresented and non-dominant communities by placing the creation of GenAI tools within the respective culture's context. If adopted effectively, these strategies promise to democratize the benefits of GenAI technology and address the diverse needs of global educational communities, ultimately ensuring that students from all backgrounds feel represented and valued within these evolving technological spaces (Calzati, 2021).

### *Foresight by Design*

Foresight by Design (FBD) is a proactive and human-centric approach to strategic planning and development, enabling developers and organizations to predict and mitigate the potential harm of GenAI while elevating diverse perspectives (Buehring & Liedtka, 2018; Dorton et al., 2023). FBD encourages a comprehensive consideration of trends, emerging technologies, social changes, and other factors that could influence the future trajectory of GenAI systems. Additionally, FBD involves a thorough assessment of the data used to train GenAI algorithms, including factors such as ethnicity, gender, socioeconomic status, and geographic location. By identifying potential biases in the training data, FBD can potentially preemptively address the perpetuation of power imbalances and oppressive outcomes through GenAI-driven decision-making and content generation.

### *Decentralizing GenAI Hubs*

Decentralizing GenAI hubs is an emerging approach to mitigating GenAI neocolonialism. GenAI's neocolonial nature creates a concentration of locations where technological innovation and development isolate the job market in Western tech hubs. Ultimately, this limits the perspectives contributing to GenAI technology development. Furthermore, opportunities to grow within the GenAI sector often require relocation for education and labor. The growing need for more jobs related to GenAI should spark debate on decentralizing the hubs from dominating countries and creating more local design locations. Examples of this decentralization include Google's GenAI lab in Accra, Ghana, and IBM's research office in Nairobi, Kenya. By diversifying the locations of GenAI development, developers can be local to where opportunities and problems are identified. Local solutions and data will be more relevant by working with local voices and within the proximal context. For example, at Google's GenAI lab, developers are improving GenAI's natural language understanding by coding roughly 2,000 languages spoken in Africa. More opportunities exist for tech hubs within countries in Africa, Asia, and Latin America, where local voices, experiences, and histories create more accurate and culturally relevant solutions in sectors like healthcare and education technology.

The development of tools that utilize GAI has the potential to revitalize and reclaim non-dominant and indigenous languages and cultures. An example is Te Hiku Media, a nonprofit Māori radio station operated by Peter-Lucas Jones and Keoni Mahelona. Jones and Mehelona's vision is to revitalize the Māori language, Tte Rreo, while still having control of the data rights. According to Hao (2019), "They overcame resource limitations to develop their own language GAI tools, and created mechanisms to collect, manage, and protect the flow of Māori data so it won't be used without the community's consent, or worse, in ways that harm its people." Together, they enlisted the local Māori community, 2,500 people, to provide verbal data to support the creation of a speech recognition model. The Māori project represents GAI development that utilizes FBD, liberatory design methods, and local context to develop a platform that serves both the people who created it as well as global users. This example of country-specific GAI use of localized GAI deployment is an approach to mitigating neocolonialism within GAI systems. GAI has become a topic for digital-territorial colonialism where retrieval and use of information are centered around Western culture (Mohamed et al., 2020). Through a colonial political economy lens, a common GAI market can be criticized since it helps maintain a dominant culture (Carmel & Paul, 2022).

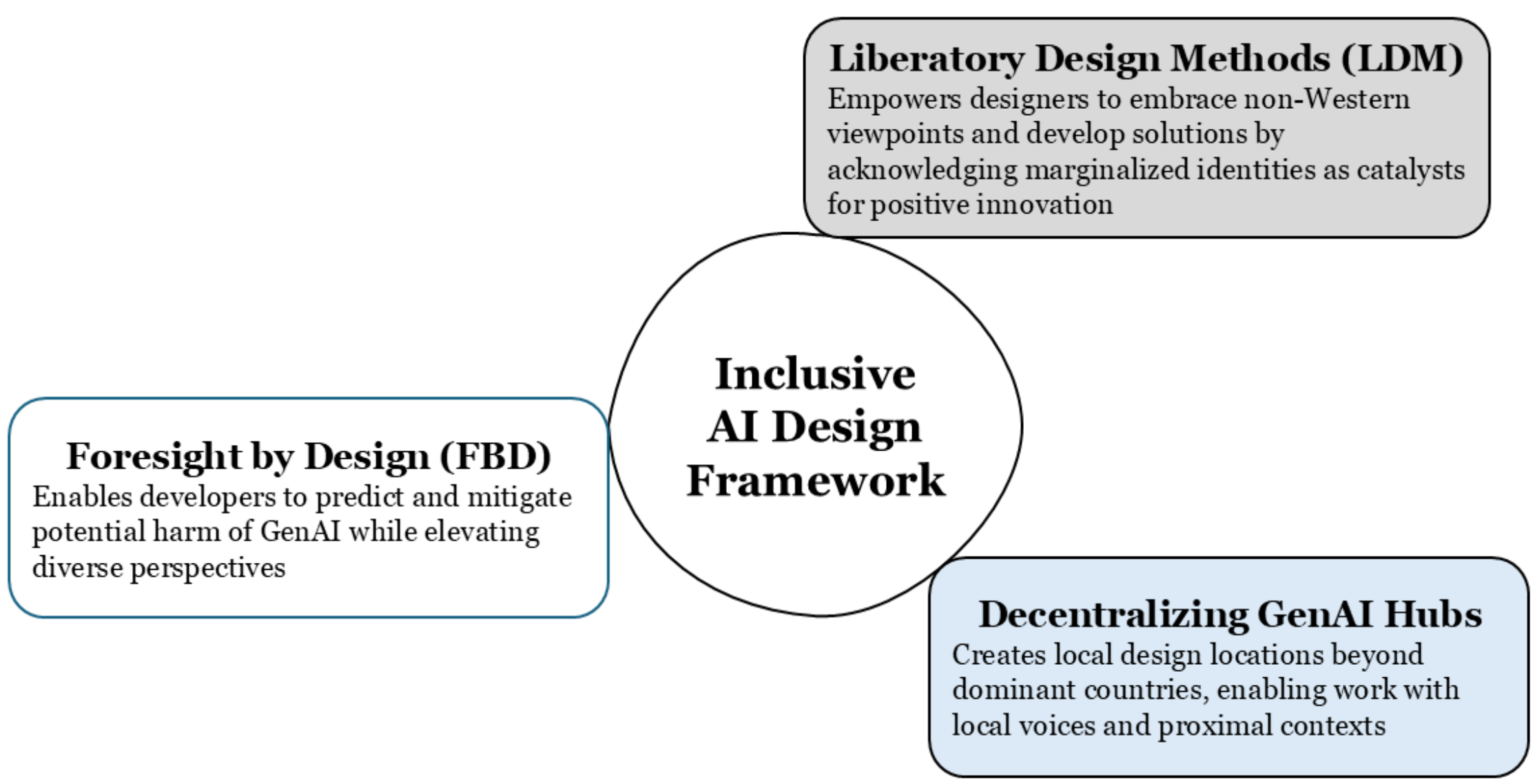

Figure 9: *Inclusive AI Design Framework*

### Human-Centric Pedagogical Level Intervention

Vulnerable groups and non-dominant populations are often marginalized and overlooked in developing educational technology (EdTech) tools and sourcing data for GenAI (Gaskins, 2023). This poses significant risks, as algorithmic bias can become embedded within these technologies, as discussed in the preceding section. To mitigate these concerns, it is important to adopt a human-centric perspective when designing and deploying GenAI in any educational setting (Karakose et al., 2023). A human-centered approach considers the end-user and the contextual environment in which the tool is utilized, while also considering whether GenAI promotes diverse cultural norms and cultural pluralism (Fishman, 2004). We therefore proposed the *Human-Centric Prompt Engineering (HCPE)* model, which positions educators and learners at the center of generative AI use in education (see Figure 10). The framework begins with contextualizing prompts to reflect local cultural settings and specific educational needs. This leads to advanced skills development, emphasizing teacher competencies and student empowerment in crafting and interpreting AI-generated content. These two primary components collectively contribute to reducing neocolonial power dynamics by minimizing forms of cultural and epistemic oppression embedded in educational materials, while also promoting equitable and culturally responsive learning through inclusive content generation (Issaka et al., 2022; Khazanchi & Khazanchi, 2024).

*Effective Promptings Strategies*

The transition from generic (such as zero-shot) to contextual prompting represents a critical skill for educators and students (Ekin, 2023). Recent investigations have established that contextualized prompting, incorporating local cultural contexts and specific educational needs, significantly enhances output relevance and inclusivity (Mollick & Mollick, 2023; Nyaaba & Zhai, 2024a). As illustrated in Figure 10, this involves two key dimensions:

1. *Contextualizing Prompts*: Embedding local cultural contexts and specific educational needs within query structures.
2. *Advanced Skills Development:* Building teacher competencies and student empowerment through systematic prompt engineering training.

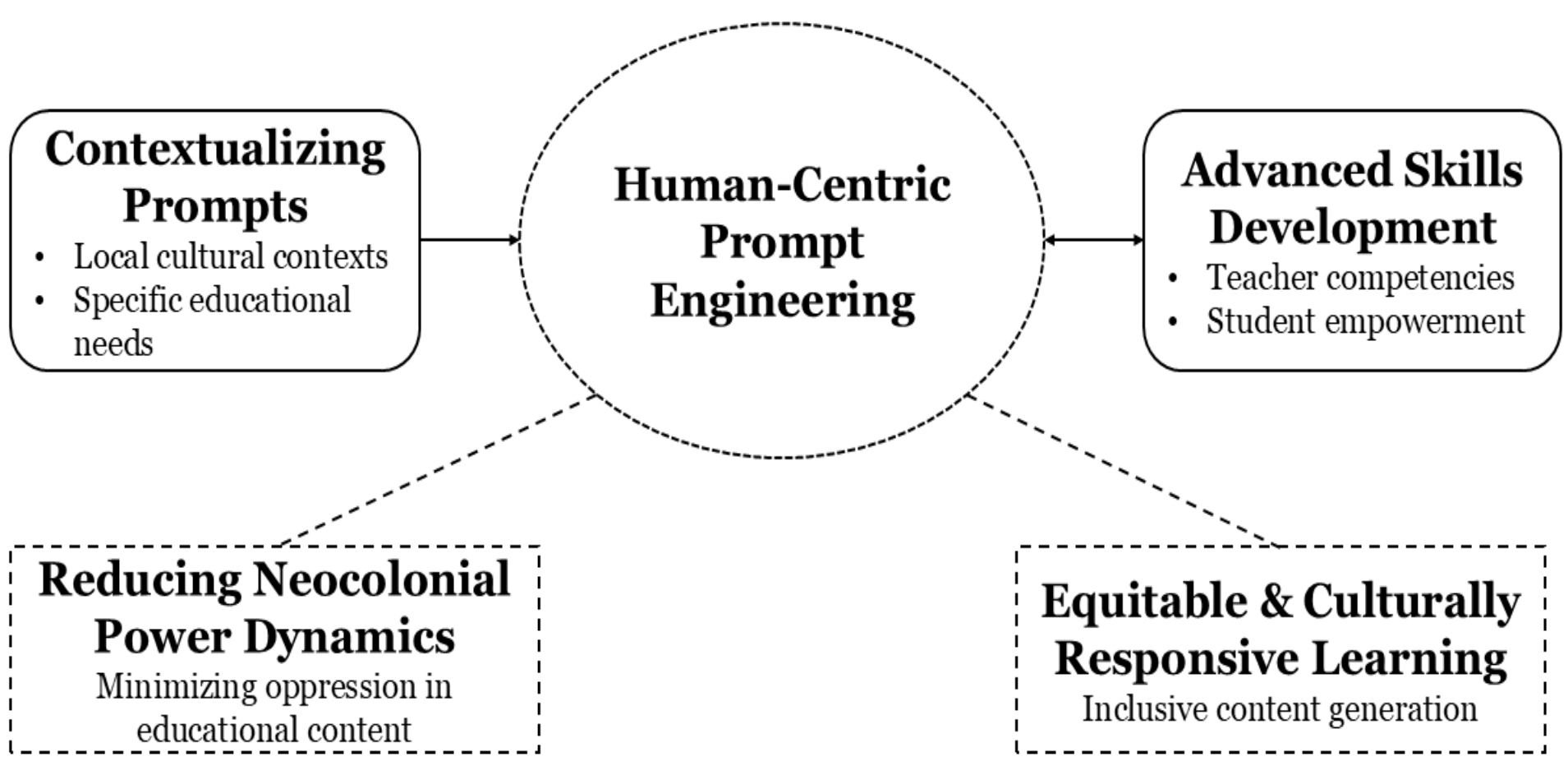

Figure 10: *Human-Centric Prompt Engineering framework*

## Implications

In this study, we examine how the biases and ethical concerns of GenAI might perpetuate neocolonial power dynamics within education. Generative AI tends to reflect Western ideologies by imposing Western values and norms on non-Western regions. This dynamic does not only risk cultural imperialism but also widens educational disparities between Western and Non-Western regions. While we encourage the use of GenAI in classroom settings to promote bias awareness, we understand that these tools cannot be effective without educators first developing competency to facilitate them. In order to use GenAI for culturally relevant teaching, teachers need competency in GenAI prompting, as well as, culturally relevant instructional strategies (Sanusi & Olaleye, 2022). Ladson-Billings (2014) posited that cultural competence is one of the factors for culturally relevant pedagogy. Educators are encouraged to strive to enhance their cultural competence through open discussions and implementing professional learnings (Sanusi & Olaleye, 2022).

We further suggest that educators harness their prompting skills in using these tools by directing prompts towards specific cultural contexts, especially as these tools are generically representing Western cultures.  The dominance of Western bias within data sources can perpetuate misinformation and invade the classroom. For example, in a situation where ideologies and images only represent whites or Western culture, students from different cultures and races might not feel belonging in these technological spaces (Obermeyer & Mullainathan). There should be a conscious effort to incorporate representation of culture, race, ethnicity, and knowledge. While this theory has been more of a suggestion for the developers of GenAI tools, we extend this suggestions to educators as well to assist students to use GenAI to identify and analyze their cultural objects, to reveal their unique qualities (Chaves & Gerosa, 2021).

Students from historically marginalized or underrepresented cultural backgrounds must be empowered to critically examine their own cultural identities and develop analytical skills to recognize the pervasive influence of Eurocentric, neocolonial norms within educational systems (Ge et al., 2024).  These suggestions depict two distinct approaches; encouraging pupils to use critical thinking skills to recognize covert colonial viewpoints in GenAI, and making use of GAI to adapt to pupils' diverse cultural backgrounds (Emenike & Plowright, 2017). A practical example may be teachers employing these GenAI tools to create a platform for discussing biases, which can be subject to critique by students. By doing so, GenAI becomes an agent in the classroom teaching students to think critically about its responses (Nayır et al., 2024).

Moreover, the perpetuation of the neocolonial aspects of GenAI may not be limited to classroom practices in education. Recent studies have shown that educational researchers largely depend on GenAI for data and many aspects of research activities such as data analysis, literature reviews, and report writing (Nyaaba et al., 2024; Owoahene Acheampong & Nyaaba, 2024). While these tools are supportive in these proposals, we suggest that educational researchers incorporate human-centric, liberatory design methods and foresight by design to address the prevailing neocolonial aspects of these tools

GenAI tools possess significant potential to positively impact education and serve as a means to bridge the digital divide (Bentley et al., 2024). However, it is imperative for the educational community to engage in a comprehensive reexamination of various aspects pertaining to GenAI, particularly those that contribute to power dynamics and marginalization. We therefore propose the integration of non-western data in GenAI tools through human-centric, liberatory design methods and foresight by design.

## Conclusions

This study critically interrogated how GenAI reproduces colonial hierarchies within education while simultaneously revealing its potential as a transformative instrument for equity and inclusion. We identified six interrelated dimensions of digital neocolonialism, curriculum ideology, linguistic exclusion, racialized representation, cultural invisibility, pedagogical control, and access inequity. These findings illuminate how GenAI systems continue to privilege Global North epistemologies and socio-technical infrastructures, thereby reproducing epistemic hierarchies in educational spaces. Yet, moving beyond critique, this study advanced a constructive response through the *Inclusive AI Design Framework* and *Human-Centric Prompt Engineering Model*, which together reframe educators and learners as co-creators of equitable AI-mediated pedagogical practices. In these frameworks, we opined that algorithmic design and prompt engineering are deeply cultural and value-laden processes. Therefore, equity in AI cannot be achieved through technical refinement alone but requires critical human agency, ethical reflexivity, and participatory inclusion across diverse educational contexts. Through contextualizing prompts, decentralizing AI development, and cultivating advanced teacher competencies, the frameworks proposed here promote a shift toward culturally responsive and inclusive AI in education.

## Future Directions

Future studies should empirically evaluate the proposed frameworks across diverse educational and linguistic contexts to test their adaptability and impact on learning equity. Longitudinal and participatory design studies could explore how teachers and students co-develop prompt strategies that resist algorithmic bias while enhancing local relevance and representation. Moreover, research must investigate AI governance, data sovereignty, and model training ethics from Global South perspectives, emphasizing how indigenous and multilingual datasets can reshape the moral architecture of GenAI systems. Advancing this agenda will require interdisciplinary collaboration among educators, computer scientists, and policymakers to ensure that the future of AI in education is intelligent and also just, anchored in cultural humility, human dignity, and global epistemic balance.

## Limitation

Although this study offers important insights into how GenAI systems reproduce and can potentially resist colonial hierarchies in education, several limitations must be acknowledged. First, the analysis was limited to two GenAI systems, ChatGPT-4 Turbo and Gemini 1.5, whose architecture, training data, and alignment processes are proprietary. Consequently, interpretations of bias and representation were derived from model outputs rather than from full transparency of internal algorithms or datasets. Second, while the study's zero-shot prompt testing provided a valuable means of comparing responses across Global North and South contexts, results are bounded by the specific prompts and contexts used. Alternative prompt formulations, languages, or domains could yield different outcomes. Third, as a qualitative design experiment, the findings emphasize depth over generalizability. Additionally, the interpretive coding of textual and visual outputs, though systematically conducted, remains influenced by researcher positionality and cultural standpoint. Future replications involving cross-cultural research teams could enhance interpretive validity and reduce subjectivity. Finally, the proposed Inclusive AI Design Framework and Human-Centric Prompt Engineering Model are conceptual and have yet to undergo extensive classroom-based validation. Empirical testing in diverse settings will be essential to refine these models, assess their pedagogical impact, and ensure their practical sustainability in addressing global AI inequities.